\documentclass[useAMS,usenatbib]{mn2e}

\usepackage{latexsym,amsmath,amsbsy,amssymb,epsfig,graphicx,wasysym}

\title[Uniting Old Stellar Systems]{Uniting 
Old Stellar Systems: From Globular Clusters to Giant
Ellipticals}

\author
[Forbes et al.]
{Duncan A. Forbes,$^{1\ast}$ Paul
Lasky,$^{1,2}$ Alister W. Graham,$^{1}$  Lee Spitler,$^{1}$\\
\\
\normalsize{$^{1}$Centre for Astrophysics and Supercomputing,}
\normalsize{Swinburne University, Hawthorn VIC 3122, Australia}\\
\normalsize{$^{2}$Centre for Stellar and Planetary Astrophysics, 
Monash University,}
\normalsize{Clayton, VIC 3800, Australia}\\
\\
\normalsize{$^\ast$To whom correspondence should be addressed; E-mail: dforbes@swin.edu.au}
}

\begin{document} 




\maketitle


\begin{abstract}

Elliptical galaxies and 
globular clusters (GCs) have traditionally
been regarded as physically distinct entities due to their
discontinuous distribution in key scaling diagrams
involving size, luminosity and
velocity dispersion. 
Recently this
distinctness has been challenged by the discovery of stellar
systems with mass intermediate between those of GCs and dwarf ellipticals
(such as Ultra Compact Dwarfs and Dwarf Galaxy Transition
Objects). 
Here we examine the relationship between the virial and
stellar mass for a range of old stellar systems, from 
GCs to giant ellipticals, and including such Intermediate Mass
Objects (IMOs). 
Improvements on previous work in this area 
include the use of (i) near-infrared magnitudes 
from the 2MASS survey, (ii) aperture corrections to velocity dispersions, 
(iii) homogeneous half light radii 
and (iv) accounting for 
the effects of non-homology in galaxies. We find a 
virial-to-stellar mass relation that ranges from $\sim$10$^4$
M$_{\odot}$ systems 
(GCs) to $\sim$10$^{11}$ M$_{\odot}$ systems 
(elliptical galaxies).   
The lack of measured velocity dispersions for dwarf ellipticals
with --16 $>$ M$_K$ $>$ --18 ($\sim$10$^8$ M$_{\odot}$)
currently inhibits our ability to determine how, or indeed if,
these galaxies connect continuously 
with GCs in terms of their virial-to-stellar
mass ratios. 
We find elliptical galaxies to have roughly equal fractions of dark and
stellar matter within a virial radius; 
only in the most massive (greater than 10$^{11}$ M$_{\odot}$) 
ellipticals does dark matter dominate the virial mass. 
Although the IMOs 
reveal slightly higher virial-to-stellar mass ratios than 
lower mass GCs, this may
simply reflect our limited understanding of their IMF (and hence their stellar
mass-to-light ratios) or structural properties. 
We argue that most of these intermediate mass
objects have similar properties to massive GCs, i.e. IMOs 
are essentially massive star clusters. 
Only the dwarf spheroidal
galaxies exhibit behaviour notably distinct from the other stellar systems
examined here, i.e. they display 
a strongly increasing virial-to-stellar mass ratio
(equivalent to higher dark matter fractions) with decreasing stellar mass.
The data used in this study is available in electronic format.

\end{abstract}

\section{Introduction}

The scalings between basic parameters such as the size,
luminosity or surface brightness, and
line-of-sight velocity dispersion 
of stellar systems have provided a key tool
in which to understand self-gravitating systems. When viewed as a
3 parameter space, they are collectively know as the Fundamental
Plane (\cite{djorgovski87}).   
Such scaling relations have
been used to probe the structural properties, origin,  
and even to classify objects depending on where they lie
in parameter space.
The scalings in `$\kappa$-space' (with axes related to mass,
mass-to-light ratio and surface brightness) of dynamically hot
galaxies were explored  
by \cite{bender92}. These hot systems included
elliptical, dwarf spheroidal and the bulges of spiral
galaxies. \cite{burstein97} extended the $\kappa$-space analysis
to include disk galaxies, groups and clusters of galaxies and
globular clusters (GCs). More recently, \cite{zaritsky06} defined
the Fundamental Manifold of spheroids revealing a continuity from
clusters of galaxies to dwarf ellipticals, and possible extension
to dwarf spheroidals.

Like elliptical galaxies, GCs are self-gravitating systems 
with a strong component of pressure
support from the random motions of their stars (i.e. they are
dynamically hot) and are dominated by stars of old age (i.e. older
than 10 Gyrs). However, they were either excluded from past
studies (e.g. \cite{bender92}; \cite{zaritsky06}) or found to be
distinct entities based on their different scalings and
large separation in mass 
from galaxies (e.g. \cite{kormendy85}; \cite{burstein97}).  

Only in the last decade have objects of mass intermediate between
those of massive GCs and dwarf ellipticals been discovered
(\cite{hilker99}; \cite{drinkwater00}). These 
objects have
masses of $\sim$ 10$^{7}$ M$_{\odot}$ and relatively 
compact sizes with measured half light radii of $\le$50 pc. 
They are usually referred to as 
Ultra Compact Dwarfs (UCDs) or Dwarf Globular Transition
Objects (DGTOs). Although they share many properties with the nuclei of
nucleated dwarf galaxies (e.g. \cite{geha03,cote06,boker08})
they also resemble massive GCs
(e.g. \cite{kisslerpatig06,hasegan05,hilker07,gilmore07}). 
A number of papers have proposed various
possible mechanisms to explain such intermediate mass objects
(IMOs; we prefer this terminology as it describes their physical
state and not their uncertain origin). 
These include the remnant
nucleus of a stripped dwarf galaxy (\cite{drinkwater03}; \cite{bekki03}), 
the merger of several smaller GCs (\cite{fellhauer02}; \cite{bekki04}), a completely 
new type of galaxy (\cite{drinkwater00}) or an extension of the GC sequence 
to higher masses (\cite{mieske02}). 
However, each of these possible explanations has difficulties
(e.g. \cite{goerdt08}; \cite{evstigneeva08}).

A number of authors have examined the scaling relations of IMOs,
sometimes including GCs and galaxies in their analysis
(\cite{martini04,hasegan05,evstigneeva07,hilker07}).  A recent
work in this fast moving field is that of 
\cite{dabringhausen08} (hereafter D08) who
include GCs and giant ellipticals, but focus on IMOs and dwarf
galaxies. They confirm a transition in globular cluster and IMO
properties (to larger sizes, higher stellar densities and higher
inferred mass-to-light ratios) at a mass of $\sim$10$^6$
M$_{\odot}$. They interpret this as either as evidence for dark
matter or a different initial mass function (IMF) 
in these somewhat higher mass objects. D08 also included dwarf
spheroidal (dSph) galaxies in their analysis. These galaxies have
similar velocity dispersions to globular clusters but very high inferred
mass-to-light ratios. Debate continues whether these high ratios 
are due to tidal heating or large dark
matter halos (see for example \cite{penarrubia08} 
and \cite{metz08}), and how such
galaxies fit into the general scaling relations. 

After submission of our paper to the journal, the work of \cite{mieske09}
was made publicly available. 
This work discusses the nature of UCDs focusing on
their internal dynamics and re-examining various UCD scaling
relations. In the appropriate sections of this paper we comment on
the \cite{mieske09} results. In general, they reach similar conclusions to us.


Here we focus on the relationship between virial and stellar
mass for a wide mass range of old, pressure-supported systems. In
particular, we examine elliptical galaxies and  globular clusters along
with IMOs and dwarf spheroidals. In general, 
such systems contain little, if any, cold or hot gas and so
the stellar mass is a good proxy for the baryonic mass in these
systems. They are usually dominated, in mass, by old stellar populations.
We also apply
several improvements on previous work through the use of:\\ 
(i) near-infrared magnitudes which are a much better
tracer of stellar mass than optical light;\\
(ii) aperture corrections to the literature velocity dispersions
of unresolved GCs and IMOs to reflect central values; \\
(iii) half light radii that account for the deviations in galaxy
light profiles from the simple R$^{1/4}$ law;
and \\
(iv) variations to the calculated virial mass for non-homology
effects between galaxies. 

In Section 2, we describe the physical parameters which we use,
while  
Section 3 lists the data samples (which are given in Table 1 and
available fully in  
electronic form).   
Section 4 presents the scaling relations of both velocity
dispersion and radius with near-infrared luminosity before examining
the virial versus stellar mass relation. Finally, in Section 5, we 
highlight prospects for future work and give our
conclusions.

\section{Data Parameters}

\subsection{Velocity Dispersions}

Luminosity-weighted central velocity dispersions ($\sigma_0$) 
from the literature have been used for 
spatially resolved objects, such
as Galactic GCs and elliptical galaxies.
However for extragalactic IMOs and GCs in the
nearby galaxies M31
and NGC 5128, spatially
resolved velocity dispersion profiles are generally not available
and hence aperture measurements are quoted in the literature. Assuming
that the velocity dispersion profiles follow the same form as
those observed for Milky Way GCs, then $\sigma_0$ can
be estimated following the 
prescription given by \cite{djorgovski97}.
Thus unless central values are quoted for IMOs and unresolved GCs in nearby
galaxies we assume a 10 percent increase in the literature
aperture velocity dispersion.  
This corresponds to a 20 percent, or 0.08 dex, increase 
in log mass. When corrections have been applied this is noted in
Table 1. 

A more sophisicated approach to the aperture
corrections of velocity dispersion measures were made by
\cite{mieske09}. Their
corrections, based on mass modelling, for the NGC 5128 GCs 
range from about 7\% to 21\%, with an average of
$\sim$ 15\%. 
If our corrections are
systematically too low by 5\% this will cause the central
velocity dispersion to be underestimated by 0.02 dex and 0.04 dex
in mass. Such small effects have no effect on our conclusions.

\subsection{Near-Infrared Magnitudes}

Rather than use optical luminosities as done
by most
previous studies (including D08), we use the K$_{\rm total}$ 2.2 
micron near-infrared magnitudes
from the homogeneous 2MASS survey (\cite{jarrett03}). This 
has several advantages: it is a better tracer of the
underlying mass as it is less influenced by young, blue stars; it
significantly reduces the influence of dust reddening; and it
reduces the sensitivity to metallicity variations from low to
high mass objects (this is important when calculating stellar
masses). 
K-band magnitudes are not corrected for Galactic
extinction as the corrections are generally negligible. 
The main disadvantage of the K-band is that fewer objects are
available for analysis. 

Our sample of elliptical galaxies (described in Section 3) 
all have total K-band magnitudes available from the 2MASS survey. A  
large subset of our galaxy sample also have total B magnitudes
available from NED.  This allows us to derive a new B--K
colour-magnitude relation for elliptical galaxies down to faint
magnitudes, with the caveat that NED photometry is very inhomogeneous. 

\begin{figure}
\begin{center}
\includegraphics[height=0.7\columnwidth,width=0.99\columnwidth]{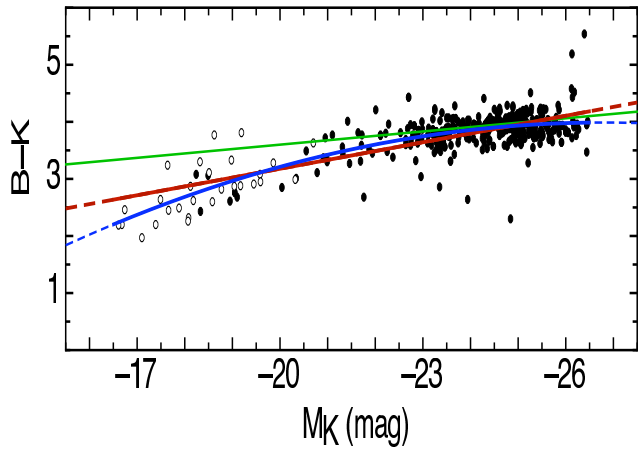}
\caption{\label{BminK} $B-K$ colour-magnitude relation. Filled circles 
are from the present dataset, with the open circles from Binggeli
\& Jerjen (1998).  The middle red and lower blue 
lines are respectively the linear and quadratic 
best fits. The upper green line is from a linear fit to Coma ellipticals by
Mobasher et al. (1986).
}  
\end{center}
\end{figure}

In Figure \ref{BminK} we show the B--K colour magnitude relation
for our elliptical galaxy sample,  
together with additional data 
from \cite{binggeli98} who provide
B-band photometry 
for a large sample of dwarf ellipticals in the Virgo
cluster. Magnitudes in the K-band for the Binggeli \& Jerjen sample 
are taken from 2MASS.
In a study of Coma cluster ellipticals, \cite{mobasher86} report a
linear fit to the B--K colour-magnitude relation for their sample
(which is plotted as the upper green line in Figure \ref{BminK}).
Whilst this fits well at the bright end, it appears to 
overestimate the colours for the dwarf elliptical galaxies.  We have fit
both a linear and quadratic function to the data
in Figure \ref{BminK}. They are:
\begin{align}
B-K = &-0.155M_{K}+0.082,\\
B-K = &-0.017{M_{K}}^{2}-0.906M_{K}-8.135.
\end{align}


The fits are statistically similar with a chi-squared of 32.6 and
29.2 for the linear and quadratic fits respectively. 
In this work we have somewhat arbitrarily chosen to use 
our linear rather than quadratic fit to the colour-magnitude
relation. Although the 
difference between the fits can be as large as $\sim$ 0.5 mag. at M$_K$ = --16
mag., the bulk of the data used in this paper (shown as filled circles in
Figure \ref{BminK}) have M$_K$ $<$ --20 and hence the difference
between our linear and quadratic fit is insignificant for our purposes.

\subsection{Half Light Radii}

For GCs and IMOs we use the half light radii as quoted in the 
original data source (see Section 3 for details). For galaxies however, 
we use the 2MASS survey to obtain half light radii. 
The 2MASS survey calculates an 
R$_{20}$ size which is the major-axis radius of the K-band 20th mag.
arcsec$^{-2}$ isophote (\cite{jarrett00}). 
We convert this 
into a half light radius (R$_h$) using an empirical 
relation based on S\'ersic light profiles, which we 
verify for a small sample
of galaxies that have both R$_{20}$ and 
R$_h$ (from S\'ersic fits to
the light profiles) independently available. 

The S\'ersic profile is given by
\begin{align}
\mu(R)=\mu_{0}+\frac{2.5b_{n}}{\ln 10}\left(\frac{R}{R_{h}}\right)^{1/n},\label{mur}
\end{align}
where $\mu_{0} \equiv \mu(R=0)$ is the central surface brightness, $n$
is the S\'ersic shape parameter and $b_{n} \approx 1.9992n-0.3171$, for
0.5 $<$ $n$ $<$ 10.  
Substituting $\mu(R_{20})=20$ into
equation (\ref{mur}) and allowing for the effect of the
ellipticity gives
\begin{align}
\frac{R_{20}}{R_{h}}=\frac{1}{\sqrt{1-\varepsilon}}\left[\frac{\ln10}{2.5 b_{n}}\left(20-\mu_{0,K}\right)\right]^{n},\label{convert}
\end{align}
where $\mu_{0,K}$ specifies the central surface brightness in the
K-band and R$_h$ is the geometric-mean  half light radius rather than
the major-axis half light radius.  
Thus equation (\ref{convert}) provides an analytic
relation, based on the S\'ersic profile, for converting
R$_{20}$ into R$_h$.

To apply equation (\ref{convert}) to our galaxy dataset we use 
the empirical relations between the S\'ersic index and central
surface brightness with galaxy magnitude.  \cite{graham03} provide
an empirical relation for the B-band absolute magnitude as a
function of $n$, such that
\begin{align}
M_{B}=-9.4\log_{10}n-14.3,\label{MBn}
\end{align}
and an empirical
relation between the central surface brightness and the absolute
magnitude, such that 
\begin{align}
M_{B}=\frac{2}{3}\mu_{0,B}-29.5.\label{MBmu}
\end{align}
To convert equations (\ref{MBn}) and (\ref{MBmu}) into K-band
magnitudes we use the B--K colour-magnitude relation (given in
equation 1).
However, when applying colours to the central surface
brightness, a correction is required due to any colour gradients
that may exist. 
\cite{michard05} discuss this in some detail, and a simple
analysis indicates
\begin{align}
\mu_{0,B}-\mu_{0,K}\approx\left(B-K\right)+0.2.\label{col}
\end{align}
Equations (\ref{convert}-\ref{col}) give an empirical relation for converting
$R_{20}$ to $R_{h}$
\begin{align}
\frac{R_{20}}{R_{h}}=\frac{1}{\sqrt{1-\varepsilon}}\left\{\frac{-3\ln 10}{5 b_{n}}\left[M_{K}+\frac{1}{3}\left(B-K\right)+16.03\right]\right\}^{n},
\end{align}
where the S\'ersic shape parameter is given by
\begin{align}
\log_{10}n=-\frac{1}{9.4}\left[M_{K}+\left(B-K\right)+14.3\right].
\end{align}
We note that for elliptical galaxies the
S\'ersic index does not vary between the B and R-bands
(\cite{graham07}) and we have assumed no variation between the
B and K-bands. Thus, equation 8 together with equations 1 and 9,
allows one to convert the 2MASS R$_{20}$ radii into half light
radii using only the 2MASS K-band magnitude and the galaxy ellipticity.

\begin{figure}
\begin{center}
\includegraphics[height=0.7\columnwidth,width=0.99\columnwidth]{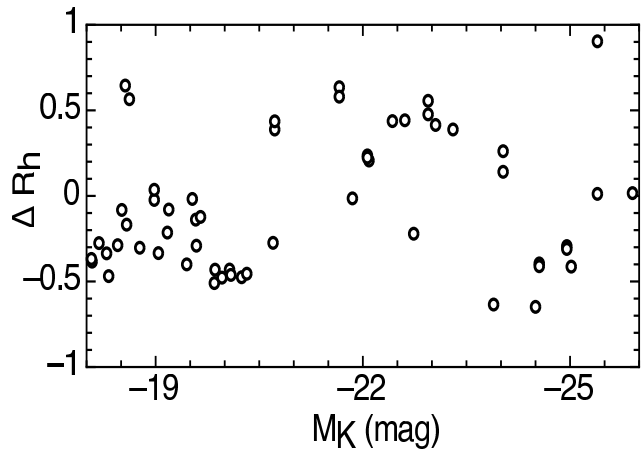}
\caption{\label{ronrvmag}		
Normalised difference between the predicted and observed 
half light radii for a range of galaxy K-band magnitudes. Here 
$\Delta$ R$_h$ equals R$_h$ (predicted) minus R$_h$ (observed) divided by 
R$_h$ (observed). 
Data come from the observational measurements of R$_h$ by 
\citet{caon93}, \citet{donofrio94} and \citet{binggeli98}. 
Although the scatter is large there is no strong 
trend with galaxy absolute magnitude.}
\end{center}
\end{figure}

To test the above procedure we use the data from
\cite{caon93}, \cite{donofrio94} and \cite{binggeli98}, who provide a sample of
elliptical galaxies with measured half light radii, 
B-band absolute magnitudes, ellipticities and surface
brightnesses.  We take R$_{20}$ and the K-band apparent
magnitudes from 2MASS for this sample of galaxies and use
equation 1 to calculate their B--K colours. This
dataset allows us to compare the observationally-derived 
R$_h$ with the expected value according to the discussion above.  
These differences, normalised by the observed R$_h$ value, 
are shown in Figure \ref{ronrvmag}. The distribution shows 
large scatter, with little systematic trend, except perhaps at low
luminosities when our predicted half light radii tend to be {\it
underestimated} by $\sim$ 40\%. The implications of this are
discussed in section 4.2.

\begin{figure}
\begin{center}
\includegraphics[height=0.7\columnwidth,width=0.99\columnwidth]{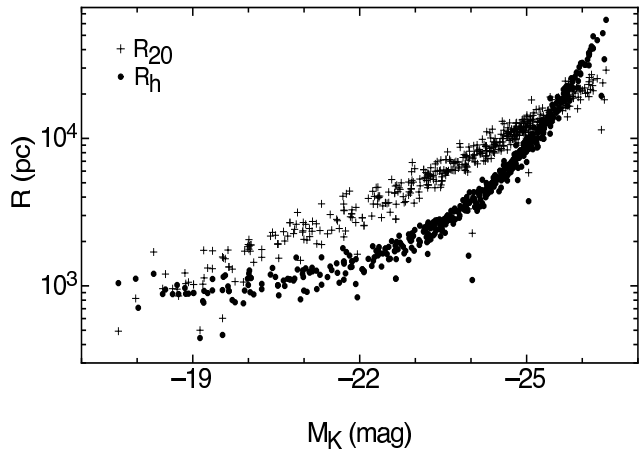}
\caption{\label{r20tore} 
Elliptical galaxy sizes versus K-band absolute magnitude 
for our galaxy sample.
The crosses represent the R$_{20}$ radii from 2MASS. The 
filled circles represent our calculated R$_h$ half light radii. 
The procedure outlined in Section 2.3 modifies the near straight
relation of R$_{20}$ with magnitude to a curved one for R$_h$. 
The correction for ellipticity, from major axis radii to
geometric-mean 
radii, has reduced the scatter in the R$_h$ relation. 
}
\end{center}
\end{figure}

Figure \ref{r20tore} shows the effect on the size-luminosity
relation for elliptical galaxies of converting 2MASS R$_{20}$
major-axis radii into R$_h$ geometric-mean radii. 
Our procedure modifies the nearly straight
relation between R$_{20}$ and magnitude to the curved trend seen
in Figure \ref{r20tore} and reduces
the scatter (the latter due to the conversion from major-axis to
geometric radii). The curved
relation is consistent with that reported by \cite{graham08} 
who did not convert from 2MASS R$_{20}$ radii but
used measured half light radii directly from S\'ersic law fits to
light profiles. We note that recent high resolution imaging with
the Advanced Camera for Surveys on HST
of early-type Virgo cluster galaxies confirms that their surface
brightness profiles are well fit by S\'ersic laws over a large
range in galaxy luminosity (\cite{ferrarese06}). 
 
We note that the error
in log R$_h$, and hence log mass, from assuming an R$^{1/4}$ law for all
galaxies can be up to a factor of 5--10 at the extreme low and high mass
ends of the galaxy sequence (see \cite{trujillo01}).  
Thus studies that assume R$^{1/4}$ law
fits for all elliptical galaxies can be subject to large systematic
errors depending on the mass range explored.

\section{Data Samples}

Here we briefly mention the literature data samples which we have used, 
in rough order of increasing mass, 
and if any specific corrections have been applied. 
We note that the classification of objects below is naturally 
somewhat arbitrary.   
Half light radii, in arcseconds, 
have been transformed into physical radii and apparent magnitudes 
into absolute ones, using 
distances from the original literature source 
for Milky Way and local volume (within 10 Mpc) objects. 
For more distant galaxies we
use the cosmic microwave background 
corrected distance with H$_0$ = 73 km s$^{-1}$ Mpc$^{-1}$ from
NED. A full compilation of the object names,
distances, half light radii, velocity dispersions and K-band
magnitudes for our sample is available electronically (see Table 1).

\subsection{Milky Way and M31 Globular Clusters}

Central velocity dispersion measurements are taken from 
\cite{mclaughlin05} and \cite{barmby07} and
references therein.
For the Milky Way GCs, no aperture correction is required and we
simply take the quoted central values. However for M31, at a distance of
780 kpc 
(\cite{barmby07}), a 1 arcsec slit corresponds to 
3-4 pc (a typical
GC half light radius) and so we increase the measured velocity 
dispersions by 10\% as described in Section 2.1. 
Effective half light radii and distances for the Milky Way GCs
come from the \cite{harris96} catalog. Half light radii for M31
GCs are taken from \cite{barmby07}. 
For M31 GCs the K-band magnitudes are taken directly from 
2MASS. For the Milky Way GCs we use the V--K colours of
\cite{cohen07}, which is based on 2MASS data 
and the total extinction corrected V-band
magnitudes of \cite{harris96}. 

Our sample includes the Milky Way globular cluster 
NGC 2419. This is a very luminous, extended GC
located in the Galaxy outer regions. Otherwise it appears to be a normal 
GC with no evidence of multiple stellar populations (\cite{ripepi07}).
We also include 037-B327 located in M31. Although heavily extincted,
\cite{cohen06} argues this is a normal GC, albeit a massive one
with a velocity dispersion of $\sim$ 20 km s$^{-1}$.

\subsection{Milky Way and M31 Possible Dwarf Galaxy Nuclei}

{\bf Omega Centauri} \\
Omega Centauri is the most
luminous GC in the Galaxy. It reveals evidence for multiple 
age and metallicity stellar
populations and has been suggested as the remnant nucleus of a
stripped dwarf galaxy (e.g. \cite{hilker00}). Here we use 
V--K = 2.27 based on
the empirical metallicity vs V--K relation of \cite{cohen07} to
calculate the K-band magnitude.\\

\noindent
{\bf M54}\\
This object is traditionally classified as
a globular cluster but is also identified as the nucleus of the
accreted Sagittarius dwarf galaxy and has multiple stellar populations (\cite{siegel07}). \\

\noindent
{\bf NGC 2808} \\
Normally classified as a GC, NGC 2808 may be the remnant nucleus of the
Canis Major dwarf galaxy (\cite{forbes04}) and it reveals multiple stellar
populations (\cite{piotto07}). \\

\noindent
{\bf G1} \\
This massive star cluster in M31 has also been suggested
as the remnant nucleus of a stripped dwarf galaxy and reveals evidence
for multiple metallicity populations (\cite{meylan01}). We have taken
the aperture velocity dispersion for G1 from
\cite{djorgovski97}, and corrected it following Section 2.1 to
give 27.6 km s$^{-1}$. In this case, a comparison can be made to
the value from an HST/STIS measurement by
\cite{gebhardt05} who derived a central velocity dispersion of
31.1 $\pm$ 1.7 km s$^{-1}$.

\subsection{NGC 5128 Massive Globular Clusters}

The velocity dispersions and half light radii for massive GCs around
NGC 5128 are taken from \cite{martini04} and \cite{rejkuba07} assuming a distance of 3.9 Mpc.
Both of these studies quote aperture velocity dispersions which
we correct to a central value following Section 2.1. 
Of the GCs in \cite{rejkuba07}, those labelled 
HGHH92 (\cite{harris92}) do
not contain K-band measurements in 2MASS.  We therefore used
the total V-band magnitude given in \cite{rejkuba07} and the $V-K$
colours given in \cite{harris92} to derive K-band
magnitudes.  All other NGC 5128 GCs were found in the 2MASS point source 
catalogue. We
include the most massive GC highlighted in the \cite{rejkuba07} study,  
i.e. HCH99-18 with a mass of $\sim$ 10$^7$ M$_{\odot}$.

\subsection{Young Massive Star Clusters}

{\bf W3 and W30}\\
W3 and W30 are 400 Myr old, massive star
clusters located in the galaxy NGC 7252 (\cite{maraston04,bastian06}). Before
including these {\it young} clusters with our old stellar systems, we evolve
them by 10 Gyrs  using the single stellar population (SSP) model
of \cite{bruzual03} and a \cite{chabrier03} IMF, which
implies a K-band fading of 2.66 mags. \\

\noindent
{\bf G114}\\
We include G114 which is a 4 Gyr old, massive star
cluster located in the galaxy NGC 1316 (\cite{goudfrooij01,bastian06}). Again
we fade the cluster (by 0.46 mag. in the K-band) to be 10 Gyrs.

\subsection{Intermediate Mass Objects (IMOs)}

Velocity dispersions and half light radii measurements for IMOs
classified as UCDs and DGTOs are from 
\cite{hasegan05}, \cite{evstigneeva07} and \cite{hilker07}.  We corrected the 
quoted aperture velocity dispersions from \cite{hasegan05} and
\cite{hilker07} (their Table 5) following Section 2.1; for \cite{evstigneeva07}
we use their estimated central value. 
The \cite{evstigneeva07} UCDs and UCD2 and UCD3 from \cite{hilker07} 
UCDs have K-band
measurements from 2MASS. The \cite{hasegan05} dataset
includes metallicities which we use to convert the quoted V magnitudes
into K using V--K colours from the SSP model of \cite{bruzual03}
and an assumed old age of 12 Gyrs. The average V--K colour of
2.84 is used to convert the V-band magnitudes of UCD4 and UCD5
from \cite{hilker07} to the K-band. 
We use a distance of 18.1 Mpc to the Virgo IMOs from
\cite{caon93} (see \cite{mei07} for a recent SBF-based distance) 
and 19.0 Mpc to the Fornax IMOs from \cite{hilker07}.

Three of the Fornax UCDs have been re-observed and their central
velocity dispersions derived by modelling their light profiles
(\cite{mieske09}). Despite our uniform 10\% aperture correction,
the values agree quite well (i.e. UCD2 27.2 vs 27.1 km s$^{-1}$,
UCD3 29.3 vs 29.5 km s$^{-1}$ and UCD4 32.1 vs 30.3 km
s$^{-1}$). Mieske et al. also presented new measurements for a
dozen additional Fornax UCDs, but unfortunately K-band magnitudes
are not available for them in the 2MASS catalogue. They also lack
age and metallicity estimates.

We also include the object M59cO. It 
is an old (9.3 Gyr), metal-rich (--0.03 dex) 
object that \cite{chilingarian08} suggest is a 
transition between a UCD and a compact elliptical-like galaxy.
Using the \cite{bruzual03} SSP 
model and \cite{chabrier03} IMF 
we estimate V--K = 3.1 and derive a K-band magnitude from
their quoted V magnitude. We use their estimated central velocity
dispersion.

\subsection{Dwarf Elliptical Nuclei}

\cite{geha02} give size, velocity dispersion and V-band magnitude
data for several nucleated dwarf ellipticals (dE,N). They also
derive properties for the nuclear component separately.
In a study of 45 dE,N galaxies, \cite{lotz04} found the nuclei
colours to corelate with global colours and be only slightly 
bluer than the rest of the galaxy (i.e. $\le$ 0.15 in
V--I). \cite{cote06} found no strong colour gradients in their
sample of dwarf ellipticals in the Virgo cluster.
Here we make the approximation that the nuclei have
the same V--K colours as observed for the entire galaxy and 
hence we calculate M$_K$ for the nuclei based on their observed
nuclei V-band magnitudes and total galaxy 
K-band magnitudes from 2MASS. We note that the use of the virial theorem 
to derive masses may not
be strictly applicable to these systems as they are not isolated,
but rather embedded within a host galaxy. 

\subsection{Dwarf Spheroidals}

Velocity dispersion measurements and V-band magnitudes for the
Local Group dwarf spheroidals are taken from 
\cite{burstein97}, \cite{mateo98}, \cite{gilmore07}, 
\cite{lewis07} and \cite{belokurov06}.  Half light radii
are from \cite{gilmore07}.  To convert V-band magnitudes into the
K-band we use V--K colours which are based on the observed
metallicity and an 
assumed old age (\cite{grebel03}) with the SSP model of \cite{bruzual03}. The 
resulting colours are V--K $\sim$ 2.

\subsection{Compact Ellipticals}

We include M32 in our sample which is 
the prototype for the class of objects called compact
ellipticals (cE). It is generally thought that M32 was
originally a much larger galaxy that has been stripped of its
outer stars hence reducing its total 
luminosity and size but with only a small 
effect on its central velocity dispersion 
(\cite{valluri05}; \cite{howley08} and references therein). 
We take the central velocity dispersion value from
\cite{vandermarel98}, the `bulge' half light radius and bulge R
band magnitude from \cite{graham02}, and R--K = 2.1 based
on matched aperture values given in NED. If we had used the total K-band
magnitude from 2MASS (which may be contaminated by light from
M31) we would derive a value for M$_K$ which is half a magnitude more
luminous. We assume a distance to M32 of 780 kpc.

\subsection{Dwarf ellipticals}

The central velocity dispersions for dwarf ellipticals come from
\cite{pedraz02}, \cite{simien02}, \cite{geha03} and
\cite{vanzee04}. 
Total K-band magnitudes and 
half light radii (converted from R$_{20}$ values) come from 2MASS.

The size and magnitude information for dwarf ellipticals is
supplemented by the Virgo cluster 
dwarf elliptical sample of \cite{binggeli98}. To the best of
our knowledge there is no velocity dispersion measurements for
most of their sample. 
The quoted B-band magnitudes were converted into the K-band using 
the linear fit to the B--K colour-magnitude relation as described in 
Section 2.2. A distance of 18.1 Mpc to the Virgo cluster dwarfs 
is used (\cite{caon93}). 

\subsection{Normal and Giant Ellipticals}


The division between normal and giant ellipticals is somewhat
arbitrary. However, a transition around M$_B$ $\sim$ --20.5 (M$_K$ $\sim$
--24.5) separates the core profile giants from the power law profile 
ellipticals of lower mass. 

We take the central velocity
dispersions of normal and giant ellipticals from \cite{bender90},
\cite{bender92}, \cite{burstein97}, \cite{faber89},
\cite{trager00}, 
\cite{moore02}, \cite{matkovic05} and \cite{firth08}.
Lenticular galaxies are excluded
from our sample. 
Total
K-band magnitudes come from 2MASS. Half light radii are
calculated from the 2MASS R$_{20}$ radii as per Section 2.2. 
We note that D08 used half light radii directly from \cite{bender92} which assumes H$_0$ = 50 km s$^{-1}$ Mpc$^{-1}$ 
and that each galaxy is well fit by an R$^{1/4}$ law (i.e. a fixed
S\'ersic n value of 4). Here we use the cosmic microwave
background corrected distances with H$_0$ = 73 km s$^{-1}$
Mpc$^{-1}$ from NED.  
We do not include the bulges of spirals from
\cite{bender92} as done by D08. Such systems may have a
substantial contribution from rotational support (which is not
taken into account by D08).

\section{Results and Discussion}

\subsection{Velocity Dispersion}

\begin{figure}
\includegraphics[height=0.7\columnwidth,width=0.99\columnwidth]{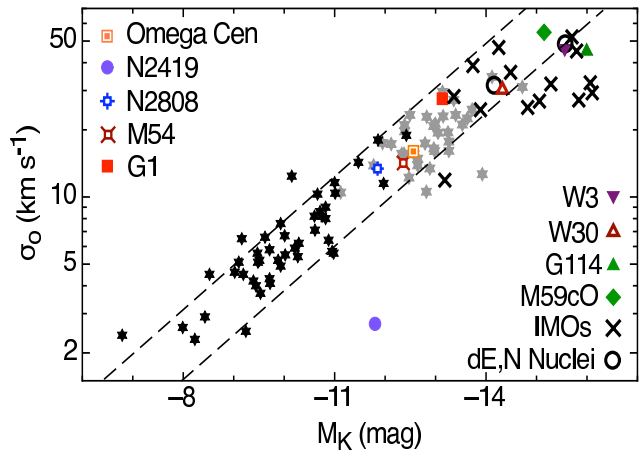}
\caption{\label{f4}Central velocity 
dispersion versus absolute K-band magnitude 
for globular  clusters and IMOs. Galactic and M31 GCs are shown by
dark stars (with NGC 2419 as a filled circle) and NGC 5128 GCs by
light stars. Possible remnants of dwarf galaxy nuclei (Omega Cen.,
M54, NGC 2808 and G1) and young massive star clusters
(W3, W30 and G114) are labelled. Intermediate mass objects (UCDs
and DGTOs) are shown by
crosses with M59cO as a diamond. dE,N nuclei are given by open circles. 
The globular clusters and IMOs follow a similar 
relation. See text for details. The dashed lines that include
most of the data are consistent with a relation
of the form L$_K$ $\sim$ $\sigma^2$. 
}
\end{figure}

Figure \ref{f4} shows that the central velocity dispersion scales
with absolute magnitude 
for objects with M$_K$ $\ge$ --16 mag. Here we show GCs, young massive star
clusters (after 10 Gyrs evolution), possible remnants of dwarf galaxy
nuclei, the nuclei of dE,N galaxies and IMOs (i.e. UCDs, DGTOs and
M59cO). We also highlight the outer Galactic GC NGC 2419 which has a
low velocity dispersion for its magnitude.  For a relation
of the type L$_{K}$ $\propto$ $\sigma_0^\beta$, the GCs have a slope
consistent with $\beta$ = 2.0, as found by \cite{mclaughlin00} in the V
band.  This is expected from the virial theorem for objects with a
constant size and M/L ratio. The plot shows that the GCs of M31 and
NGC 5128 follow the general trend of Galactic GCs to brighter
magnitudes.  This supports the conclusion of \cite{barmby07} that
GCs in these galaxies share common structural properties. The proposed
remnants of dwarf galaxy nuclei (Omega Cen, M54, NGC 2808 and G1)
occupy the high luminosity (mass) end of the GC sequence but are also
consistent with the trend for GCs. The half dozen most luminous
IMOs (i.e. M$_K$ $\le$ --15 mag.) show some evidence for smaller
velocity dispersions at a given K-band magnitude. However, more data are needed
to verify if this is a systematic trend.

\begin{figure}
\includegraphics[height=0.7\columnwidth,width=0.99\columnwidth]{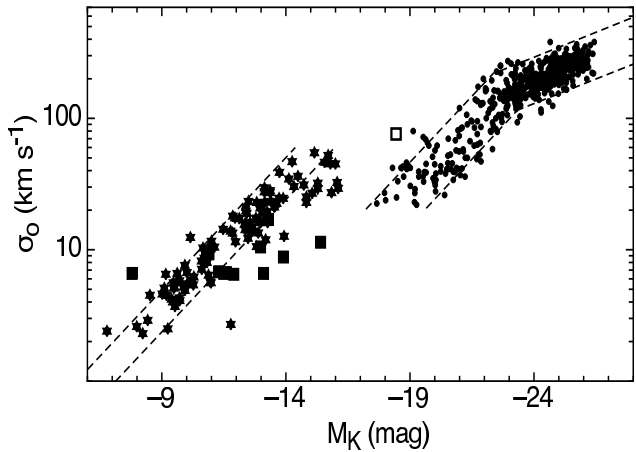}
\caption{\label{f5}Central velocity dispersion versus absolute K-band magnitude.
Star clusters and IMOs from Figure \ref{f4} are shown as stars,
elliptical galaxies as filled circles and dSph
galaxies as filled squares. The cE galaxy M32 is shown as an open
square. The dashed lines roughly indicate the parameter space occupied by
the locus of points. For a relation of the type 
L$_K$ $\sim$ $\sigma_0^{\beta}$ the slopes  
($\beta$) of the dashed lines are 4 
for giant ellipticals, and 2 for the dwarf
ellipticals and globular clusters/IMOs.  
The dSph galaxies are offset from both the globular cluster and
galaxy relations.  }
\end{figure}

We note that most low luminosity dE galaxies contain nuclei
(\cite{binggeli98}; \cite{graham03}; \cite{cote06}),
and so may actually be classified as dE,N in good quality
imaging. Thus to varying degrees, measurement of the velocity
dispersion in such galaxies 
is being influenced by the presence of the nucleus. In the
case of the Local Group dE,N galaxy NGC 205 the nucleus is `cold' and
the velocity dispersion profile reveals a prominent dip at the galaxy
centre (\cite{valluri05}; \cite{howley08}). However, the velocity dispersion
profiles for a small sample of dE,N galaxies by \cite{geha02} reveals a
range of profiles from central dips to peaks.  
Geha et al. derived an estimate of the nuclear properties including
the velocity dispersion. These nuclear velocity dispersions and
magnitudes are consistent with the massive GC and IMO trend (see
Figure \ref{f4}).

Figure \ref{f5} extends the central velocity dispersion versus
absolute magnitude plot to
include elliptical galaxies.  Galaxies follow a 
L$_{K}$ $\propto$ $\sigma_0^\beta$ relation that has 
$\beta$ $\sim$ 4 for
giant ellipticals (the well known Faber-Jackson relation) which
changes to $\beta$  $\sim$ 2
for dwarf ellipticals (e.g. de \cite{derijcke05}; \cite{matkovic05}). 
Thus extrapolation of the Faber-Jackson relation for giant
ellipticals down to the luminosity of GCs and IMOs, as done in
some studies, would seem inappropriate. 
We note that dwarf elliptical slope 
is the same as that for GCs, albeit with an offset in
the relations of $\Delta$M$_K$ $\sim$ 5 mag. between low luminosity
dwarfs and massive GCs. 
As has been long recognised, the rare cE galaxy M32 
does not lie on either the giant or dwarf luminosity-$\sigma$ sequences.

We also show the location of dSph galaxies in Figure \ref{f5}. They have a
range of absolute magnitudes that span from GCs to the most luminous IMOs,
however they possess near constant velocity dispersions. 
The
evolution of dark matter dominated dSph galaxies under the influence
of tidal stripping has been explored by \cite{penarrubia08}. They
find that tidal stripping reduces the stellar luminosity, radius and
central velocity dispersion while increasing the mass-to-light
ratio. Their tidal evolutionary tracks can qualitatively explain the
properties of the dSph galaxies, however the dSph galaxies in Figure \ref{f5}
have a shallower trend than predicted by their models. An
alternative scenario is that they are dynamically evolved tidal
dwarf galaxies (\cite{metz08}).

\begin{figure}
\includegraphics[height=0.7\columnwidth,width=0.99\columnwidth]{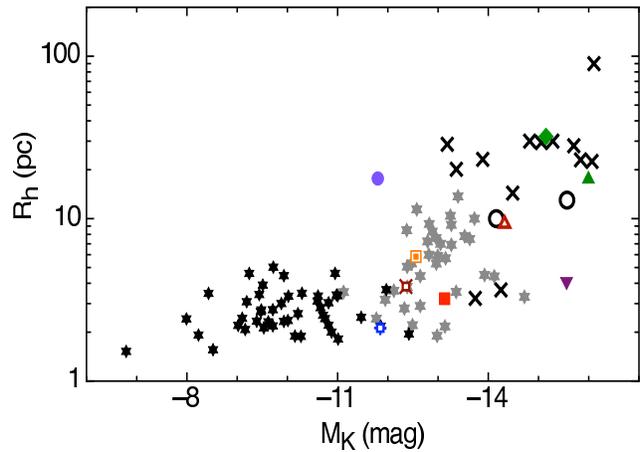}
\caption{\label{f6}Half light radius versus absolute K-band magnitude 
for star clusters and IMOs. Symbols as in Figure \ref{f4}. Globular clusters 
with M$_K$ $>$ --12 reveal a near constant size, while more luminous 
globular clusters and IMOs scatter about a linear luminosity-size relation.
}
\end{figure}

\subsection{Half Light Radii}

Figure \ref{f6} 
shows the half light radius, R$_h$, versus absolute K-band
magnitude for objects with M$_K$ $\ge$ --16 mag.
GCs with 
a luminosity of M$_{K}$ $\ge$ --12 mag. (which corresponds to a
stellar mass of $\sim$ 10$^6$ M$_{\odot}$) 
have a near constant size of 3-4 pc. 
More luminous GCs, young massive star clusters, remnants of dwarf galaxy
nuclei, the nuclei of dE,N galaxies and IMOs, have sizes that scale roughly 
linearly with luminosity (e.g. \cite{kisslerpatig06}). 
This may also be interpreted as a lower
envelope to the distribution of half light radii with magnitude (see
\cite{barmby07}). The outer Galactic GC NGC 2419 has
a very large size for its magnitude. We note that the so-called 
`faint fuzzy' objects (\cite{brodie02}) have larger sizes 
than typical low mass GCs but are not included in this analysis as they 
lack internal velocity dispersion measurements.

\begin{figure}
\includegraphics[height=0.7\columnwidth,width=0.99\columnwidth]{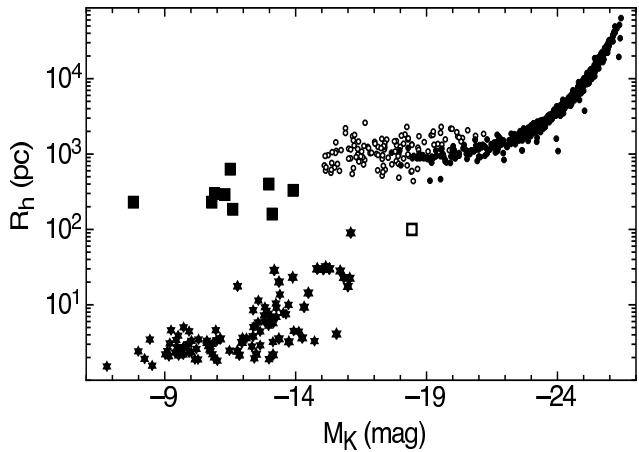}
\caption{\label{f7}Half light radius versus absolute K-band magnitude. 
Symbols as in Figure \ref{f5}, except dwarf ellipticals from
Binggeli \& Jerjen (1998) have been included as open circles. These low luminosity dwarf
galaxies indicate that the galaxy size-luminosity sequence
flattens to a near constant size. The (near empty)
gap in size of 50-100 pc appears to be a real feature. dSph
galaxies do not follow the general galaxy size-luminosity
relation. 
}
\end{figure}

Figure \ref{f7}  shows the half light radius versus absolute
magnitude 
including
elliptical and dSph galaxies. 
The giant ellipticals have 
a linear slope similar to that 
found by \cite{fish64}, this relation flattens for lower
luminosity ellipticals to a roughly constant half light radius of 1 kpc. The
dSph galaxies have a range of half light radii, which is
generally smaller than the lowest luminosity dwarf elliptical. 

D08 also examined the luminosity-size relation and to quote them
{\it ``It is surprising that the MCOs [IMOs] lie on the same relation
between mass and radius as massive elliptical galaxies with
masses $\ge$ 10$^{11}$ M$_{\odot}$, while elliptical galaxies with
lower masses (i.e. objects in the intermediate mass range) mostly
lie on a different relation, which points towards the parameter
space of dSphs.'' } 

Figure \ref{f7} does indeed show that giant ellipticals and 
massive GCs/IMOs have a similar luminosity-size relation slope 
However, we consider this as coincidental as  
the elliptical galaxy relation extends continously from giants 
to dwarfs with no obvious break. In other words, no single linear
luminosity-radius relation is suitable for elliptical 
galaxies over a wide range of luminosity.  

As for the dSph galaxies, Figure \ref{f7} indicates that their half
light sizes are smaller than a simple extrapolation of the dwarf
elliptical relation to lower masses.  
We note that D08 have used data directly
from \cite{bender92} which assumes sizes based on H$_0$ = 50 km
s$^{-1}$ Mpc$^{-1}$ and that R$^{1/4}$ models are good fits to the
light profiles of dwarf galaxies (which they are not).  Here we
have used H$_0$ = 73 km s$^{-1}$ Mpc$^{-1}$ and calculated half
light radii from S\'ersic models (using equations 8 and 9).  In order
to illustrate the dwarf sequence to lower luminosities we include the
half light radii mured by \cite{binggeli98} who fit S\'ersic
models to their data. The
K-band absolute magnitudes are determined from a linear fit to our B--K
colour-magnitude relation (Section 2.2),so the uncertainty in the
B--K colours at M$_K$ = --16 may cause a $\sim$0.5
mag. change in the \cite{binggeli98} 
M$_K$ values plotted, i.e. a horizontal shift
only.  We note that the \cite{binggeli98} data scatters evenly
about an extrapolation of our sample to lower luminosities. 
This indicates that our half light radii conversion from 
2MASS R$_{20}$ radii is reasonable even at low luminosities. 
We conclude that dE sizes remain roughly constant at a value
some 5-10$\times$ greater than the typical size of dSph galaxies.
The possible systematic {\it underestimate} of our half light
radii by 40\%, or 0.15 dex at M$_K$ $\sim$ --19 (see Figure \ref{ronrvmag})
would only serve to increase the gap in sizes between low
luminosity dEs and dSph galaxies.   
The possibility that larger size dSph galaxies exist, but have
yet to be discovered, can not be ruled out. 

The general lack of
objects with sizes around 50-100 pc appears to be a real feature
rather than a selection effect. The tidally stripped galaxy M32 and UCD 
number 3 from the study of \cite{hilker07}
are notable exceptions. 
\cite{gilmore07} 
have argued that the gap represents
a physical divide between dark matter {\it free} star clusters and IMOs, and 
the dark matter {\it dominated} galaxies. To address this, and
related issues, we investigate the stellar and total virial masses for
our sample below.

\begin{figure}
\includegraphics[height=0.7\columnwidth,width=0.99\columnwidth]{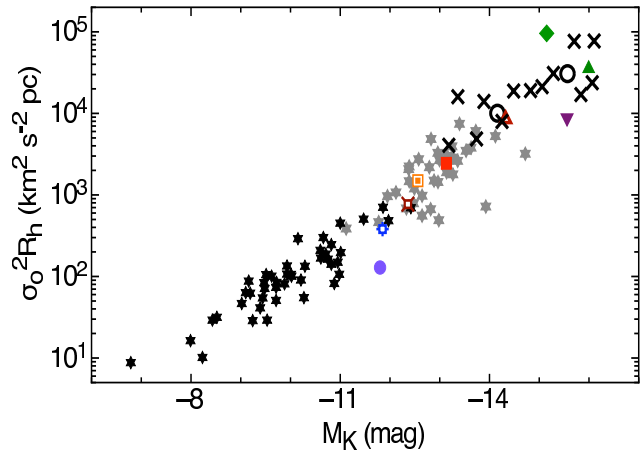}
\caption{\label{f8}Dynamical mass versus absolute K-band magnitude 
for globular clusters (GCs) and intermediate mass objects
(IMOs). Symbols as in Figure \ref{f4}. A continuous 
trend is seen from low mass GCs to massive GCs and IMOs. 
}
\end{figure}

\subsection{Masses}

Figure \ref{f8} 
shows $\sigma_0^2$ $\times$ R$_h$ (a measure of dynamical mass) 
versus absolute K-band magnitude. This figure  
reveals a continuous relation including GCs, 
young massive star clusters, remnants of dwarf galaxy
nuclei, the nuclei of dE,N galaxies and IMOs. 
The `wiggles' seen in Figures \ref{f4} and \ref{f6} have largely been
removed by considering $\sigma_0^2$ $\times$ R$_h$. Similarly, the GC NGC 2419
with a large size and low velocity dispersion is no longer an
extreme outlier. {\cite{ripepi07}} note that the half light radius in 
\cite{harris96} may be underestimated, and hence the data point for 
NGC 2419 may move upward in Figure \ref{f8} to join the general trend.

\begin{figure}
\includegraphics[height=0.7\columnwidth,width=0.99\columnwidth]{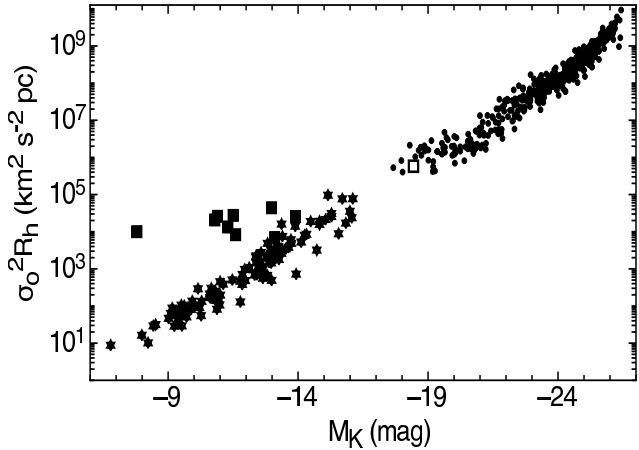}
\caption{\label{f9}Dynamical mass versus absolute K-band magnitude.
Symbols as in Figure \ref{f5}. dSph galaxies do not follow the general 
GC/IMO and galaxy trend.
}
\end{figure}

Figure \ref{f9} extends Figure \ref{f8} to include elliptical and dSph
galaxies. A general sequence is seen over $\sim$ 18 magnitudes in
M$_K$ from GCs to giant ellipticals (including the 
cE galaxy M32), with a notable gap in the
data at --16 $>$ M$_K$ $>$ --18. 
This gap can be filled by dwarf
galaxies (e.g. from the \cite{binggeli98} sample) 
when their velocity dispersions are known. If the 
L$_{K}$ $\propto$ $\sigma_0^2$ dE galaxy scaling holds then such galaxies,
with constant half light radii of about 1 kpc, will extend the
galaxy trend in Figure \ref{f9} 
down in mass to join the massive GCs and IMOs. If however,
lower luminosity dwarf galaxies reveal relatively constant central velocity
dispersions of $\sim$ 30 km s$^{-1}$ then they will not join up
with the GC/IMO trend but rather have a constant
dynamical mass (a horizontal line at $\sim 10^6$ km$^2$ s$^{-2}$
pc in Figure \ref{f9}). 
We note that although dSph galaxies have similar $\sigma_0^2$
$\times$ R$_h$ values to the IMOs, they have a range of
luminosities and hence occupy a different sequence to the other
objects.


\begin{figure}
\includegraphics[height=0.7\columnwidth,width=0.99\columnwidth]{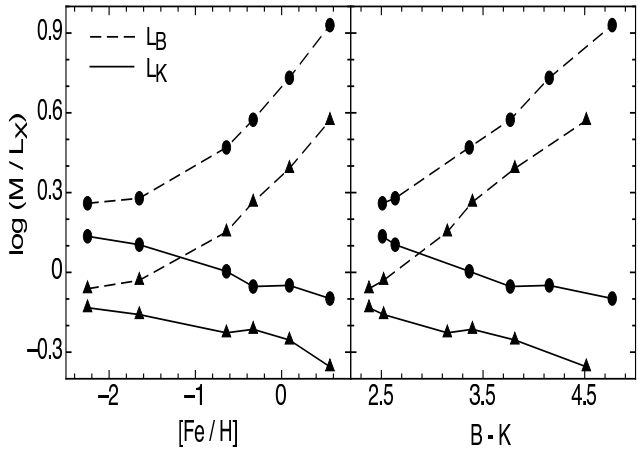}
\caption{\label{mlk}Mass-to-light ratio in the B and K-bands as a function
of metallicity and B--K colour. A linear interpolation is shown
between the single stellar population model points of Bruzual \&
Charlot (2003) using a Chabrier (2003) initial mass function for
a 12 Gyr (circles) and 5 Gyr (triangles) old population. The
mass-to-light ratio shows a smaller variation with metalliciity and B--K
color in the K-band than the B-band. 
}
\end{figure}

\begin{figure}
\includegraphics[height=0.7\columnwidth,width=0.99\columnwidth]{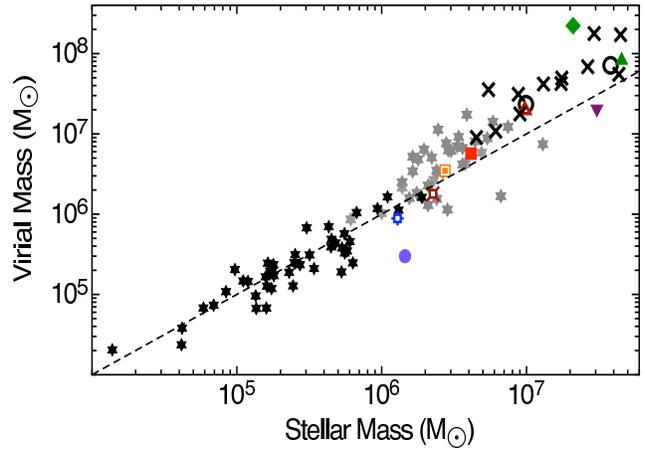}
\caption{\label{f10}Virial mass versus total stellar mass for globular 
clusters and IMOs.
Symbols as in Figure \ref{f5}. A virial coefficient of 10 is used.  
The stellar mass is calculated from the K-band magnitude 
assuming a M/L$_K$ ratio (see
text for details). A one-to-one relation is shown as a dashed 
line. Massive GCs and IMOs deviate to higher virial-to-stellar mass ratios. 
}
\end{figure}

\begin{figure}
\includegraphics[height=0.7\columnwidth,width=0.99\columnwidth]{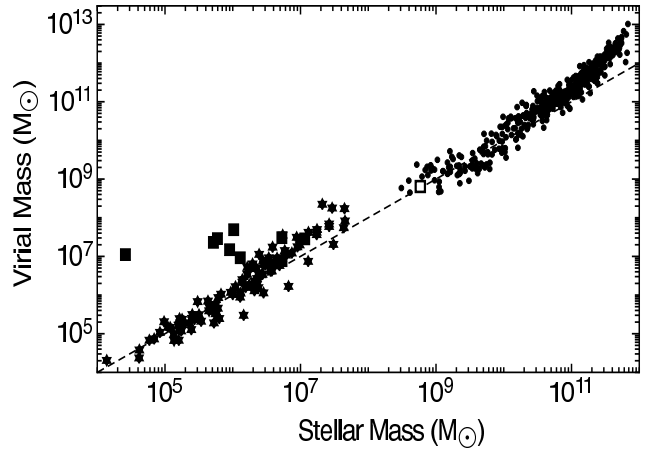}
\caption{\label{f11}Virial mass versus total stellar mass. Symbols as in Figure 
\ref{f5}.  
The dashed line represents a
one-to-one relation. A virial coefficient of 5 has been
used for the galaxies and 10 for the GCs and IMOs. 
dSph galaxies (squares, with a virial coefficient of 10) 
do not follow the globular cluster
or galaxy
sequences.
}
\end{figure}

\subsubsection{Virial and Stellar Masses}

We now probe the relationship between virial and stellar mass.
The total stellar mass, in solar masses, 
is calculated from the absolute K-band magnitude using M$_{K
\odot}$ = +3.28 and a K-band  
stellar mass-to-light ratio. We use the K-band
stellar M/L ratios from the stellar population models of \cite{bruzual03}
with a \cite{chabrier03} IMF and an assumed age of 12 Gyr. We
note that the Chabrier IMF is essentially identical to the
\cite{kroupa02} IMF, which is a good representation of the mass
function of resolved young star clusters.
Figure \ref{mlk} shows the M/L$_K$ variation as a function of
metallicity and B--K colour for an old (12 Gyr) and intermediate
age (5 Gyr) stellar system.  We have derived the best fit
relation for M/L$_K$ over the metallicity/colour range from 
metal-poor GCs
to giant red ellipticals.
Thus for a given observed colour, or 
metallicity, we assign a unique stellar M/L$_K$ to the object in
question. This effectively removes any metallicity differences
between objects and allows us to compare the resulting 
stellar masses. 
One advantage of using the K-band is that 
M/L$_K$  is a weak function of metallicity and colour 
(and hence system mass) for old stellar systems, 
varying by only $\sim$ 0.2 dex for our sample. 
Although shifted to lower M/L ratios by a factor of
about two relative to the 12 Gyr track, the 5 Gyr old
track reveals a similar variation with metallicity and colour. 
For either old or intermediate-aged systems, the M/L variations in the K-band
are substantially less than that in the B-band. 
Most previous work has used 
optical wavelengths for which the M/L ratio correction
can vary by $\sim$ 0.5 dex with 
metallicity or age and hence adds a significant source of
uncertainty in any stellar mass estimate (see discussion by D08).

To calculate stellar masses we have assumed an age of 12 Gyrs
when deriving the K-band 
M/L ratio.  
Galactic GCs are uniformly older
than 10 Gyrs (\cite{deangeli05}) and the globular clusters with 
multiple stellar populations appear to be dominated by the oldest population 
(e.g. \cite{piotto07,siegel07}). 
Direct age measurements of IMOs are very rare in the
literature, however their colours and limited 
spectra are generally consistent with old ages
(\cite{hasegan05,evstigneeva07}). Dwarf ellipticals reveal a
range of {\it central} ages (\cite{caldwell03}). 
If these systems were {\it dominated} by a
younger (e.g. 5 Gyr) population, then the M/L$_K$ ratio would be
lower by a factor of $\sim$2, and hence the log stellar mass lower by
$\sim$0.3 dex.
Although some giant ellipticals
reveal evidence for young stars at their cores, global
averages typically support very old ages (\cite{proctor08}). 
We refer the reader to D08 for an extensive discussion on
alternative stellar population models and IMFs.

The virial mass 
is calculated as $C$ $\times$ 
$\sigma_0^2$ $\times$ R$_h$ / $G$ and is an estimator of total
mass within the virial radius 
(e.g. \cite{prugniel97}). Here $G$ is the universal
gravitational constant and $C$ 
is the 
virial coefficient  which incorporates various factors such as
degree of virialisation, conversion of light-weighted into
mass-weighted quantities, projection effects, stellar orbits, etc.
See \cite{zibetti02} and \cite{trujillo04} for an alternative approaches. 

In Figure \ref{f10} 
we show the virial mass against the total stellar mass 
for globular clusters and IMOs. 
A virial coefficient of $C$ = 10 is used for these objects. It has
the property 
that the virial mass is roughly equal to the stellar mass for the
low mass GCs, which are thought to be free of dark matter today
(\cite{moore96}; Brodie 2008, priv. comm.) but may have
originally formed in dark matter minihalos
(\cite{mashchenko05}). 
We note that \cite{hasegan05} has calculated dynamical masses for
their DGTOs assuming their surface
brightness profiles are well fit by King laws. For the six DGTOs in
common, we find differences to our virial masses of between
1\% and 26\%. Thus at the extreme, the $\sim$0.1 dex difference
in mass is within the scatter of Figure \ref{f10}.

Globular clusters, young 
star clusters and IMOs have an upper mass limit in Figure \ref{f10} of about
10$^8$ M$_{\odot}$. This may be a physical limit associated with local
conditions within a galaxy, i.e. a peak in 
the star formation rate per unit area
(\cite{larsen00}) 
and the internal galaxy pressure (\cite{billett02}).
There is also a size-of-sample effect so that the galaxies with
the most star clusters host the highest mass clusters (\cite{whitmore00}). 
For such galaxies, star clusters are observed in the local
universe to have an upper limit of $\sim$ 10$^8$ M$_{\odot}$ (see
compilation by \cite{whitmore00}). To date the known IMOs are
within this upper mass limit for star clusters. This is
consistent with the idea that IMOs are massive star clusters.

Figure \ref{f10} also shows that the high mass GCs and IMOs start to
deviate systematically 
from a one-to-one line to slightly higher virial-to-stellar mass ratios 
(equivalently higher inferred M/L ratios) assuming the same
virial coefficient of 10 is appropriate. 
For the high mass GCs the two-body relaxation time is longer than 
their inferred age (D08). This fact may be
reflected in the
appearance of a luminosity-radius relation for massive GCs 
(Figure \ref{f6}), a change in slope of the luminosity-$\sigma_0$ relation 
(Figure \ref{f4}), the presence of multiple stellar populations
(\cite{piotto07}) and extended light profiles
(\cite{mclaughlin08}). 
A similar change in the virial-to-stellar mass ratio
at a few times 10$^6$ M$_{\odot}$ is found by \cite{mieske09}.

For the massive globular clusters and IMOs, 
the ratio of virial-to-stellar mass is $\sim$3:1. 
However the dark matter fraction of 2/3 suggested by this relation 
is at odds with our understanding of massive GCs as being dark matter free 
like their low mass counterparts
(\cite{moore96}; Brodie 2008, priv. comm.). This can be
reconciled for the massive GCs (and IMOs) if the virial masses
are {\it overestimated} or if the stellar masses {\it
underestimated} by a factor of about three (0.48 in log mass). 
\cite{kouwenhoven08} have found that the presence of binaries can
lead to an overestimation of the virial mass by up to a factor of
two, however this effect is strongest for low mass GCs and
largely insignificant for star clusters of mass $\sim$ 10$^7$ M$_{\odot}$.
\cite{kundu08} has recently suggested that the current 
samples of massive GCs and IMOs
are preferentially drawn from large galactocentric radii and
are hence intrinsically larger, as seen in the Milky Way
(\cite{vandenbergh91}). We have tested this hypothesis on the 16 GCs in
NGC 5128 from \cite{rejkuba07} which have a mean size of R$_h$ =
5.9 pc for a mean projected galactocentric distance of 9.4
kpc. Using the scaling method of \cite{barmby07} we find that these
GCs are some 50 per cent larger than equivalent Milky Way
GCs. Thus this selection effects may lead to an overestimate of
the virial mass (assuming an unchanged velocity dispersion) 
of $\sim$ 0.18 dex in the
log, which is insufficient to account for the trend observed.   
Another possibility is if the
massive GCs and IMOs are not fully relaxed as suggested by D08,
then using their measured velocity dispersions may overestimate
the virial mass.  
The alternative approach is to explore reasons why the stellar
mass may be underestimated.  For example, we could 
appeal to a different IMF than used here, e.g. one with 
a higher stellar mass-to-light
ratio such as a Salpeter IMF.  
This which would increase the
stellar mass (by about 0.2 dex in log mass). However, there is
little observational support for a Salpeter-like IMF in
resolvable star clusters (\cite{chabrier03}). 


Figure \ref{f11} extends Figure \ref{f10} 
to include elliptical and dwarf spheroidal galaxies. 
As seen in Figure \ref{f9}, 
Figure \ref{f11} reveals a gap in virial mass at a few 10$^8$ to 10$^9$
M$_{\odot}$.  
Dwarf spheroidal galaxies have near constant virial masses of $\sim$ 10$^7$
M$_{\odot}$ and do not appear to follow the GC/IMO or galaxy sequences. 
Elliptical galaxies are 
offset from the one-to-one relation to higher virial-to-stellar 
mass ratios. 
Below we explore the effects of using more
realistic virial coefficients for elliptical galaxies. We
initially set the virial coefficient $C$ = 5 following 
\cite{cappellari06}. In their dynamical study
of several normal ellipticals they 
concluded {\it ``... that the simple virial mass estimate of M/L,
and correspondingly of galaxy mass, is virtually unbiased, in the
sense that it produces estimates that follow a nearly one-to-one
correlation with the M/L computed from much more `expensive'
dynamical models.''}

\subsubsection{Correcting for the Effects of Non-Homology in
Elliptical Galaxies}

\begin{figure}
\includegraphics[height=0.7\columnwidth,width=0.99\columnwidth]{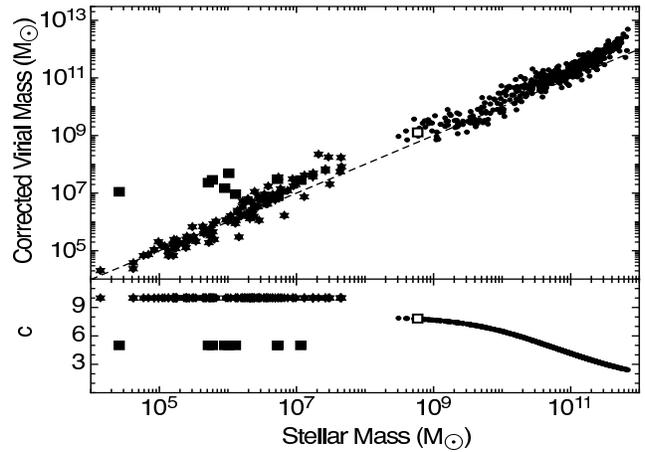}
\caption{\label{f12}Corrected 
virial mass versus stellar mass with a variable virial
coefficient for galaxies. The lower panel shows the variation of the virial
coefficient $C$ with galaxy stellar mass (from Bertin
et al. 2002). Symbols as in Figure \ref{f5}.  
The galaxy sequence has been largely straightened out compared to
Figure \ref{f11}. 
}
\end{figure}

Figure \ref{f11} assumes a virial coefficient $C$ that is the
same for all galaxies (i.e. homology). 
However, elliptical galaxies have luminosity-dependent light
profile shapes (e.g. \cite{caon93}; \cite{ferrarese06}; \cite{ball08}). 
This structural non-homology implies an associated
luminosity-dependent, dynamical non-homology.
Parameterisation by a S\'ersic law has revealed
that 
profiles vary from n $\sim$ 0.5  to 10 (\cite{caon93}).
The associated dynamical non-homology in the velocity
dispersion profile influences the tilt of the fundamental plane (FP).  If
homology effects are left uncorrected, they cause the standard 
FP to depart from the virial plane which may result in 
erroneous mass-to-light 
trends with mass (\cite{capelato95}; \cite{graham97}).  
In this paper we take 
non-homology effects into account using the virial coefficient from
\cite{bertin02} (see also \cite{prugniel97}). 
They created a spherical, non-rotating, isotropic
model galaxy with various light profile shapes.  
Their resulting virial
coefficient is appropriate for use with the half light radius and a
velocity dispersion measured within R$_e$/8 (which is similar to the
central values given in the literature). 
Using their eq. 11 which relates the virial
coefficient to the S\'ersic index n, and the known relationship of n with
magnitude (equation 5), we can effectively correct the virial mass of
galaxies for the effects of non-R$^{1/4}$ light profiles.
The result of applying this
correction 
is shown in
Figure \ref{f12}, along with the variation of $C$ with stellar
mass.
We note that for those 
normal elliptical galaxies, for which the R$^{1/4}$ model is a 
reasonable fit to the light profiles, $C$ $\sim$ 5
(\cite{cappellari06}).    
Relative to Figure \ref{f11} (which has $C$ = 10), the galaxy data have moved
0.3 dex closer to the one-to-one line.   
Giant ellipticals with a small virial coefficient ($C$ $\le$ 3) 
have been corrected downwards the most in this plot. 
The scatter about the one-to-one line is $\pm$0.3 dex, however 
we note the caveat from 
\cite{mamon05}, that even if the velocity dispersion profile were
known to high accuracy out to 5 half light radii, unconstrained
mass and stellar anisotropy profiles could make the total mass
uncertain by a systematic factor of 3 (0.48 dex in log mass).
We further caution that if there is a conspiracy between the
radial distribution of stars and dark matter, such that their
combined density profile is isothermal irrespective of galaxy
mass (\cite{bolton08}), 
then the virial coefficient should also be constant with
galaxy mass.

After accounting for 
non-homology, we find that over the K-band
(B-band) magnitude interval $-20.5 > M_{K} > -25.5$ ($-18 > M_{B} >
-21.5$), our elliptical galaxy sample 
appear consistent with the expectation of the
virial theorem.  That is, after allowing for the
primary effects which introduce a ``tilt'' into the regular FP,
our results are in agreement with the virial plane.  Figure \ref{f12}
reveals that the virial-to-stellar mass ratio does not strongly vary
with galaxy mass for masses less than about 10$^{11}$
M$_{\odot}$. A fit to the elliptical galaxy data
below masses of 10$^{11}$
M$_{\odot}$ is fully consistent with a slope of unity,
i.e. the virial mass scales directly with the stellar mass, with
the ratio of virial-to-stellar mass being $\sim$2:1 on average. 
This suggests that about half 
of the mass in dwarf to normal elliptical galaxies is in
the form of dark matter. A similar conclusion was reached by
\cite{derijcke05} in a dynamical study of dwarf ellipticals. 
The lack of variation in the ratio over
this mass regime suggests that claims of a varying
mass-to-light ratio, e.g. $M/L
\propto L^{\alpha}$ (with a
non-zero value of $\alpha$) derived from a standard FP analysis
with R$^{1/4}$ light profile fits, should
be reconsidered.

For the most massive (above 10$^{11}$ M$_{\odot}$)
elliptical galaxies the slope of the relation becomes steeper
than unity. 
Could this effect be due to the presence of an additional
baryonic component, i.e. hot gas, 
that is not accounted for in our stellar mass estimates?
Most galaxies we have considered have little if any cold or
hot gas, so that the stellar mass is essentially the same as the
baryon mass content. However, the main exception to this is that
the most massive
ellipticals may contain a substantial halo of hot gas (\cite{osullivan01}). 
\cite{matsushita01} has shown that the hot gas
content in ellipticals, as traced by X-rays, within 4 half light radii
is typically 0.1\% of the stellar mass for a 10$^{11}$ M$_{\odot}$
elliptical rising to 10\% for a 10$^{12}$ M$_{\odot}$ one. Thus
corrections for the hot gas mass in giant ellipticals would tend to
flatten the slope but this effect is not strong enough to reduce
the slope to unity, i.e. the most massive ellipticals have higher
virial-to-baryon mass ratios than lower mass galaxies.
Thus the dark matter fraction appears to increase
in the most massive ellipticals. This is
consistent with results of \cite{gallazzi06} who find evidence
for higher virial-to-stellar mass ratios in a sample of 26 000
galaxies from the Sloan Digital Sky Survey. It is also supported
by strong lensing studies (\cite{ferreras05}), 
which have probed the enclosed mass out to 5 half light
radii, finding the virial-to-stellar mass ratio varies from about 
1:1 in 10$^{10}$ M$_{\odot}$ ellipticals 
to 5:1 in 10$^{12}$ M$_{\odot}$ galaxies. We note that {\it weak}
lensing results tend to find higher virial-to-stellar mass
ratios (e.g. \cite{mandelbaum06}). 

The Bertin et al. model assumes a non-rotating galaxy, however lower
mass  
ellipticals are known to have a larger contribution to
their dynamics from rotation than high luminosity
ellipticals (e.g. \cite{bender92}). 
Ignoring rotation
and flattening tends to underestimate the virial mass and hence
the resulting M/L ratios. For (rotating) normal ellipticals,
\cite{bender92} estimate the effect is up to 0.13 dex in
log virial mass. 
From the samples of the lowest luminosity dwarf
ellipticals (M$_V$ $\sim$ --16 mag.) studied to date
(e.g. \cite{pedraz02}; \cite{geha02}; \cite{vanzee04}),
some show hints of rotation while others show none.
For giant ellipticals,
the contribution from rotation is minor. Thus the correction due
to rotation appears to be small for most elliptical galaxies (see
also \cite{matkovic05}) and 
we do not include it here. We remind the reader that 
our sample consists of only elliptical
galaxies and does not include obvious 
lenticular galaxies.

\subsection{Trends with Velocity Dispersion}

\begin{figure}
\includegraphics[height=0.7\columnwidth,width=0.99\columnwidth]{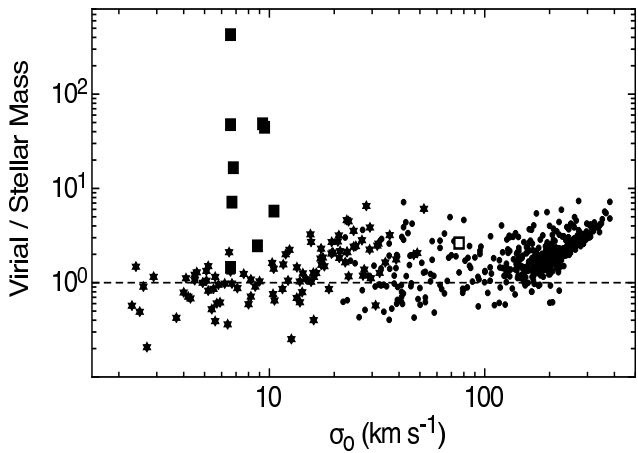}
\caption{\label{f13}
Virial-to-stellar mass ratio versus velocity dispersion. Symbols
as in Figure \ref{f12} with the dashed line showing a one-to-one
relation. Most objects are consistent with a virial mass equal to
the stellar mass, the exceptions are the most massive ellipticals
which show ratios increasing up to a factor of ten and dSph
galaxies which have ratios up to one thousand. We do not find evidence
of the `U-shape' \citep{zaritsky06}.}
\end{figure}

\cite{zaritsky06} first present the unification
of spheroidal stellar systems in what they refer to as the 
Fundamental Manifold of spheroids.   
They argue for the existence of a continuous `U-shaped'
trend between M/L and $\sigma$, with large M/L ratios for
both low mass dwarf spheroidal and high mass 
giant elliptical galaxies and clusters of galaxies. They do
not include IMOs or GCs in their analysis. 

In Figure \ref{f13} we show a similar plot to \cite{zaritsky06} 
which includes GCs and IMOs but excludes systems of galaxies,
i.e. groups and clusters of galaxies. It shows virial-to-stellar
mass ratios that scatter about a ratio of 1 (for low mass GCs)
and $\sim$2 up to $\sigma_0$ $\sim$ 200 km s$^{-1}$ at which
point the ratio increases rapidly for the most massive ellipticals. Dwarf
spheroidals appear distinct in Figure \ref{f13}  
with a large
range in their virial-to-stellar mass ratios for relatively
similar velocity dispersions. Figure \ref{f13} is qualitatively 
similar to figure 9 of Zaritsky et al. for systems with velocity
dispersions greater than $\sim$ 30 km s$^{-1}$. 
However GCs, and some IMOs, 
are inconsistent with the \cite{zaritsky06} unification which 
have relatively constant virial-to-stellar mass 
ratios, of roughly unity, as their velocity dispersions
decrease. 
Neither the simple U-shaped quadratic form as advocated by
\cite{zaritsky06}, or the revised Fundamental Manifold 
of \cite{zaritsky07}, 
provides a good representation for our sample which includes GCs
and IMOs. 

\cite{mieske09} also discuss whether GCs and UCDs fit the Zaritsky et
al. Fundamental Manifold, reaching a similar conclusion to us
that GCs deviate from the manifold.

\section{Conclusions and Future Work}

We have collected various data samples from the literature for old, 
pressure-supported 
systems which includes globular clusters, massive star clusters, 
intermediate mass objects (such as ultra compact dwarfs), 
dwarf spheroidals, dwarf ellipticals 
and giant ellipticals, and 
covers a range in mass from $\sim$10$^4$ to 10$^{12}$
M$_{\odot}$. We have applied several improvements on past work
that has examined their virial and stellar masses. 
We have employed aperture corrections to the velocity dispersion 
measurements of GCs and
intermediate mass objects. We have also derived new half light
radii for elliptical galaxies based on their sizes and ellipticities  
from the homogeneous 2MASS survey. Near-infrared magnitudes from the 2MASS 
survey are converted into total stellar masses using a K-band mass-to-light 
ratio that depends on the metallicity or colour of the object. 
Virial masses are calculated taking into account non-homology effects 
for galaxies. 


Although the scalings of velocity dispersion and half light radius
with absolute K-band magnitude vary depending on the mass regime
probed, these scalings combine to give a virial versus stellar mass
relation that shows a remarkable near continuous trend from GCs to
ellipticals. We find that the Fundamental Manifold of
\cite{zaritsky06} is not a good representation for GCs. 
Dwarf and normal elliptical galaxies are found to have 
virial-to-stellar mass ratios of $\sim$2:1.  
This ratio only increases in the very most massive
ellipticals, with masses greater than 10$^{11}$
M$_{\odot}$. 
Our results are subject to systematic effects from
remaining uncertainties, e.g. in the distribution of dark matter,
the accuracy of 2MASS total K-band magnitudes, 
the appropriate IMF to use etc.
However, such trends are generally consistent with results from strong
lensing studies.


The recently discovered intermediate mass ($\sim$ 10$^{7}$
M$_{\odot}$) objects, e.g. Ultra Compact Dwarfs and Dwarf
Globular Transition Objects, cover the same parameter space of
velocity dispersion, half light radius and mass as massive
globular clusters, possible dwarf galaxy nuclei, massive star
clusters and the nuclei of dE,N galaxies.
To date, these intermediate mass objects do not exceed the maximum
mass of known star clusters in galaxies (\cite{whitmore00}) and they are
spatially concentrated near large galaxies (\cite{wehner07}). 
All of these facts would support an interpretation that
intermediate mass objects are essentially massive star
clusters.  Given that there is no evidence in the literature 
for dark matter in massive
GCs, this would also argue against dark matter in intermediate mass
objects as they occupy a similar parameter space.  
However a mystery remains, in that intermediate mass objects
(and massive GCs) exhibit higher virial-to-stellar mass ratios,
when we apply the same virial coefficient and Chabrier IMF as low
mass GCs.  
Possible solutions to this mystery may include 
a different virial coefficient due to a longer relaxation
timescale in these systems, or 
that a bottom heavy Salpeter-like IMF is
more appropriate in these objects. 

In general agreement with D08, 
we find dwarf spheroidal galaxies to be distinct in terms of their
scaling parameters, following neither an obvious extension
of the elliptical 
galaxy or globular cluster relations. Their virial-to-stellar
mass ratios reach one thousand.   

Although we have probed mass scales from $\sim$10$^4$ to 10$^{12}$
M$_{\odot}$, there is a mass regime which remains largely unexplored
observationally, i.e. $\sim$10$^8$ M$_{\odot}$. This is greater than
the most massive star clusters and intermediate mass objects 
known but less than the mass
of dwarf ellipticals for which velocity dispersions are available. It
is not clear if the virial versus stellar mass relation will
connect smoothly across this mass gap between dwarf ellipticals
and globular clusters.  
Thus an observational campaign to
measure central velocity dispersions for a sample of very low mass dwarf
ellipticals is needed.  
Observations are also needed to determine whether the 
radial velocity dispersion profiles in intermediate mass objects
are flat or fall with radius, like their light profiles. 
Obtaining both high spectral and spatial resolution for such
small, low surface brightness systems will be observationally
challenging. 
Near-infrared 
spectra that could constrain the IMF would also be useful (\cite{mieske08}). 

On the theoretical side, for massive globular clusters and
intermediate mass objects we require a detailed understanding of how the
virial coefficient varies with the mass and/or type of stellar
system, and 
the expected 
stellar mass-to-light ratio 
that includes the effects of multiple stellar 
populations and dynamical evolutionary processes. 

\section{Acknowldegements}

We thank R. Proctor, J. Hurley, M. Hilker and P. Kroupa  
for useful 
discussions. We also thank the referee for a careful reading of
the paper and several useful comments. 
DF thanks the Faculty of ICT and the ARC for financial 
support. AG thanks the Faculty of ICT and Swinburne University
for their financial support. This research has made use of the
NASA/IPAC Extragalactic Database (NED). This research made use of
the Two Micron All Sky Survey (2MASS).




\bibliographystyle{mn2e}

\bibliography{ucd4}

\clearpage

\newpage


\maketitle

{\bf Table 1. Data Parameters}\\
{\bf Milky Way and M31 Globular Clusters}\\

\begin{center}
\begin{tabular}{lccccc}

\hline\hline
Name & Distance & $\sigma_{0}$ & $R_{h}$ &  $K$ & Reference\\
& (Mpc) & (km s$^{-1}$) & (arcsec) & (mag.)  &\\
\hline
       NGC 104  &   0.005 &   11.5  & 167.4 &  1.30 &   1, 2 \\
      NGC 1851  &   0.012 &   10.4  &  31.2 &  4.39 &   1, 2 \\
      NGC 1904  &   0.013 &    5.2  &  48.0 &  5.67 &   1, 2 \\
      NGC 3201  &   0.005 &    5.2  & 160.8 &  3.98 &   1, 2 \\
	...     &    .... &   ....  &  ...  &   ... &   ...\\
\hline
\end{tabular}
\end{center}
An asterisk implies the velocity dispersion has been corrected to
a central value following Section 2.  References are for the
velocity dispersion and the half light radii respectively.  They
are 1. Mc~Laughlin \& van der Marel~(2005), 2. Harris~(1996),

\end{document}